Regulation of Heart Beats by the Autonomous Nervous System in Health and Disease: Point-Process-Theory based Models and Simulation [V-I]

Bosco Emmanuel

Institute of Mathematical Sciences, Chennai-600113, India

**Abstract**

We have advanced a point-process based framework for the regulation of heart beats by the autonomous nervous system and analyzed the model with and without feedback. The model without feedback was found amenable to several analytical results that help develop an intuition about the way the heart interacts with the nervous system. However, in reality, feedback, baroreflex and chemoreflex controls are important to model healthy and unhealthy scenarios for the heart. Based on the Hurst exponent as an index of health of the heart we show how the state of the nervous system may tune it in health and disease. Monte Carlo simulation is used to generate RR interval series of the Electrocardiogram (ECG) for different sympathetic and parasympathetic nerve excitations.

**Introduction**

It will not be an overstatement to say that the heart is the center of human physiology. Besides providing the life sustaining fluid the blood to each and every cell of the human body, the heart constantly receives vital information from almost every other part of the body via the blood and the nervous system. These are the channels of the multi-organ communication which keep the internal body states balanced in good health and leads to catastrophic failures in disease [Journal of Cardiology 59(2012) 117-122]. Information is coded in the blood by its chemical or bio-chemical composition or its pressure. There are baroreceptors to sense the blood pressure and to inform the brain-stem by the afferent nerves and the efferent nerves pass on this neural signals via sympathetic and para-sympathetic nerves to the heart for the necessary corrective action on the SA node of the heart [**J. Math. Biol. 36(1997) 41—63**, **J. Math. Biol. 51(2005) 508-526**] for example. Then there is a large body of chemical messengers in the blood and their chemoreceptors in the body starting from the simple CO2 molecule, to which the lungs and their ventilation rates respond, to the hormones which regulate the emotional states of the subject [*Science 197(1977) 287-289*].

As is well known the basic life-sustaining process is the ion-transfer across the cell membrane and the generation and propagation of the action potentials in the nervous system and all excitable muscles. Ultimately these action potentials and the spike trains which consist of a time-ordered sequence of the action potentials codify all the vital information crisscrossing the human body [The New Cognitive Neurosciences, 3rd edition, Editor: M. Gazzaniga, MIT Press, 2004]. It is also useful to classify the ultimate drivers of these spike trains as internal to the body and as external to the body. All stimuli entering through the senses [e.g. visual, auditory etc.] are processed in the brain and an appropriate response to these external stimuli consists in the generation of spike-trains going down the nervous system and targeting specific organs of the body[*Nature Neuroscience 7(2004) 456-461, Neural Computation 14(2001) 325-346*]. Several sub-systems such as the Hypothalamus and Endocrine system play their vital roles in the regulation of these processes which govern the behavior and the psycho-physiology of the individual. Apart from such external stimuli, there are a large number of internal stimuli generated in every organ of the body and transmitted across [using the blood and the nervous system] to the target organs. Some of these stimuli are involuntary [i.e. not involving any conscious element] and others proceed from a voluntary act.

Happy and unhappy situations in life affect the nervous system: danger, mental or physical stress, human conflicts, and sickness, happy times spent in a family picnic and in a nice garden. Some of these are "fight or flight" situations and others are "rest and digest" scenarios. The happy situations strengthens the parasympathetic activity while the stressful or unhappy situations enhances the sympathetic nerve firing rates. There may be external as well as internal stimuli to the nervous system.

The present paper pertains to the influence of the nervous system on the heart. Though the central nervous system [CNS] which consists of the brain and the spinal cord and the peripheral nervous system [PNS] further divided into the autonomous nervous system [ANS], somatic and enteric subsystems constitute a hierarchy of interconnected neural network with several levels at which neural information transfer may take place, we propose to deal only with the sympathetic and parasympathetic branches of the ANS[**Journal of Cardiology 59(2012) 117-122; Neurol Neurosurg Psychiatry 70(2001) 305–310**] . The justification is two-fold: It is well known experimentally that it is the ANS which controls the heart [**Muscle

**Nerve, 36(2007) 595– 614]** besides the intrinsic cardiac controls. This is not to deny the role of the rest of the nervous system as the original recipients of the stimuli [both internal and external] that are spread across the entire nervous system. What one is saying here is that after information transfer/processing at several higher levels of the nervous system, the ANS receives this end-level information which is communicated to the heart. As a matter of fact, it could be a challenging task for neuro-scientists to trace back the origins of this end-level information at the ANS. Indeed this end-level neural information may have multiple origins both inside of the human body and the external world as perceived by the individual. Then there is the further question of how these neural signals from multiple origins superpose at the ANS level. Needless to say that we do not take up this bigger neurological challenge here. Instead we take up the similar, though more modest, challenge of understanding how the end-level neural activations at the ANS and the intrinsic cardiac activations superpose at the heart to determine the heart dynamics. However the model presented here and its possible generalizations can be a good starting point for higher levels of information processing in the nervous system. Naturally the starting point of the model will be the two neural spike trains one in the sympathetic and the other in the parasympathetic branches of the ANS and the third sequence coming from the intrinsic heart dynamics involving the SA node which is the primary pace-maker of the heart. Based on this model we advance an analytical as well as a Monte Carlo framework for studying neural influences on the heart and to generate synthetic or model ECG's. The heart has been modeled as a set of non-linear oscillators and based on such models synthetic ECG's have been reported[*IEEE TRANSACTIONS ON BIOMEDICAL ENGINEERING, VOL. 50, NO. 3, MARCH 2003]* . These oscillator models, though they possess rich physics, do not correspond to the physiology that admits only first-order-in-time dynamics and not the oscillator's second-order-in-time law. Interestingly oscillations in physiology are produced by non-linear first-order-in-time equations well known right from the days of Hodgkin-Huxley model for ion-transfer across cell membranes. The present work does not have any overlap with these earlier oscillator models and hence we do not discuss them any further.

The object of the present study is the ECG which every one of us would have seen and it is normally a standard 12-lead ECG having 12 [potential versus time] traces. These are essentially projections of the dynamic cardiac electrical vector [**Advanced Methods and Tools for ECG Data Analysis, Gari D. Clifford, Francisco Azuaje and Patrick E. McSharry (Editors), ARTTECH House, Boston, 2006]** along 12 different directions connecting the center of the heart to the 12 different lead positions. Lead

II is of common interest. If one looks closely at the trace number II, for example, one sees a series of heart beats. The potential versus time profile of each heart beat looks like a sequence of 5 amplitude variations denoted as P, Q, R, S and T.

This PQRST wave form is very rich in its information content about the heart and the cardiac function which is being routinely extracted by soft wares that come with ECG machines and is of much diagnostic value [**Computer Methods and Programs in Bio-medicine, 2 7 ( 2 0 1 6 ) 144–164**]. However our interest lies in the series of the R-events and not in this PQRST wave form per se. This R-wave is the major high-amplitude cardiac event that corresponds to the contraction of the ventricles resulting in the pumping of blood to the entire body. To be more specific the data relevant to the present study is the series of RR intervals.

While the PQRST waveform possess much information intrinsic to the heart, the RR-interval series is rich in information reaching the heart from sources extrinsic to the heart such as the sympathetic and parasympathetic nerves which are in turn fed by baroreceptors, chemoreceptors and a host of sensory and non-sensory organs in the body. The readers are directed to the net for nice illustrations showing sympathetic and parasympathetic nervous systems and their connectivity to almost every organ of the body including the heart.

Neuronal signals are carried along the nerves by action potentials in the form of a spike-train and at the synaptic cleft between the nerve terminal and the SA node of the heart, neurotransmitters are released that help trigger a heartbeat [noradrenaline for sympathetic innervation and acetylcholine during parasympathetic innervation[**Mathematical and Computer Modelling 44(2006) 952-962** ]. One Ach reaction kinetics cycle takes 15 microseconds and an Ach channel remains open for 10 m sec on the average [*NATURE | VOL 410 | 8 MARCH 2001| 277-284*]. For the SA node to trigger a heart beat a voltage threshold needs to be crossed and the time taken to cross the threshold is different for sympathetic, intrinsic and parasympathetic innervations. [**Frontiers in Physiology 3(2012) 1-9**; *Current Cardiology Reviews, 9(2013) 82-96*]. It is further known that the threshold crossing is a Wiener process (i.e., integrated white noise) leading to inverse Gaussian for the first passage time distribution **[IEEE TRANSACTIONS ON BIOMEDICAL ENGINEERING, VOL. 47, NO. 9, SEPTEMBER 2000].**

One more aspect of the heart function which should be incorporated in any framework is the refractory periods of the heart. This is a brief period of rest for the heart after every heart beat and there are several kinds of refractory periods. In this paper only one effective refractory period will be used. Besides the time for threshold crossing and the cardiac refractory period, there are also time delays for neural signals from baroreceptors and chemoreceptors to reach the SA node. The propagation of action potentials in the form of spike trains along a neuron constitutes an electrical activity and on the arrival of spikes at the synaptic cleft between the neuron terminal and the SA node the production of neurotransmitters such as Ach is initiated leading to the firing of SA node. One reaction cycle of Ach production and consumption lasts for 15 microseconds [**Mathematical and Computer Modelling 44(2006) 952-962**]. *Fraction of Ach channels open up to time t is given by exp(-lambda x t) where lambda = 0.1 per m sec [**NATURE | VOL 410 | 8 MARCH 2001 277-284**].* See the net for nice pictures of a neuron and a synaptic cleft. This exposition will be incomplete without a mention of two other physiologic waves besides the ECG. These are the blood pressure wave and the respiratory wave which directly influence and are influenced by the heart dynamics.

**Background and Prior Research**

Heart Rate Variability [HRV] is the theme of the present paper. The actors that control it are: ANS with its baroreceptor and chemoreceptor origins, the BP and the Respiratory systems, time delays in signal transmission over nerves, neuro transmitters like Acetyl Choline and Norepinephrine, internal and external stimuli like emotional stress, fear and exercise. Before we present a point-process-based framework for HRV, a quick review of the relevant earlier work is in order.

A large number of indices have been developed and employed to quantify HRV. Here we are not providing an exhaustive summary of these markers and only we point to some important earlier works.

*Time-Domain indices: SDNN in m sec: Standard deviation of all N-N intervals (from the entire recording). RMSSD in m sec: Root mean square of successive differences between N-N intervals. Higher the SDNN and RMSSD higher is the parasympathetic tone and healthier the heart [Circulation. 129(2014) 2100–2110].*

NN50 Count: the mean number of times an hour in which the change in successive normal sinus (NN) intervals exceeds 50 m sec. This is used to assess parasympathetic activity from 24 hour ECG recordings and to compare healthy subjects with those with diabetes mellitus and patients after cardiac transplantation. pNN50 statistic: defined as NN50 count/total NN count. p stands for percentage. It provides diagnostic and prognostic information in a wide range of conditions [Heart 88(2002) 378–380].

*Frequency-Domain:* [Circulation. 129(2014) 2100–2110] NLF in %: normalized low frequency content and reflects sympathetic modulation. LF band is between 0.04 and 0.15 Hz. NHF in %: normalized high frequency content reflects RSA and parasympathetic modulation. HF band is between 0.15 and 0.4 Hz.

ULF: cycle length of >5-min and ≤24-hour. Circadian and neuroendocrine rhythms. ULF band is below 0.003Hz.

VLF: effect of the renin-angiotensin system. Higher values are believed to reflect better autonomic function.

LF/HF ratio: May reflect relative sympathetic-parasympathetic activity.

[Circulation Research 59(1986) 178-193] is a classic paper reporting that the power spectral density of R-R interval variability contained two major components in power, a high frequency at ~0.25 Hz and a low frequency at ~0.1 Hz, with a normalized low frequency: high frequency ratio of 3.6 ± 0.7. Beta-adrenergic receptor blockade (0.6 mg/kg propranolol) in Man and IV nitroglycerin in Dogs were studied. Bilateral stellectomy in Dogs indicated that sympathetic nerves to the heart are instrumental in the genesis of low-frequency oscillations in R-R interval.

*Complexity indices:*

*Multiscale entropy [MSE]:* [**Phys.Rev.Lett. 89(2002) 068102-1, Phys.Rev.Lett. 92(2004)089803-1, Physical Review E 71((2005) 021906, Entropy 17(2015) 3110-3123**] MSE has recently been used to monitor critically ill patients whose life continuation relies on extracorporeal membrane oxygenator (ECMO) [**SCIENTIFIC REPORTS 5(2015) 8836**]. In **Scientific Reports 5(2015)17552**, area under MSE is used to predict Stroke-In-Evolution in ICU patients. SDNN, RMSSD, LF, HF and LF/HF ratio are measured besides the MSE area. Adjusted odds ratio (OR) of each of these parameters in predicting stroke-in-evolution are computed and tabulated.

*Fractal Measures:*

There are measures based on fractal scaling of cardiac inter beat interval dynamics [**Circulation 100 (1999) 393-399**]. Young healthy subjects had a Hurst exponent of nearly 1 and elderly and unhealthy subjects are marked by much lower Hurst exponent [**Am. J. Physiol. 271 (Regulatory Integrative Camp. Physiol. 40): R1078-R1084, 1996**].

Our hearts are not precision clocks. If it were that would point to serious cardiac disease. An optimal HRV points to a healthy heart. Human hearts beat 60 to 100 times a minute. Heart rate and life expectancy depends on the body mass: ~ 40 bpm for horses and ~ 1000 bpm for birds [*Cardiovascular Research 45 (2000) 177–184*]. Life expectancy decreases with incresing heart rate [**JACC 30**(1997) 1104 – 6]. For example, mouse with a heart rate of 600 bpm lives only for a couple of years whereas whale with less than 20 bpm lives for more than 30 years. Man is an exception: with nearly 70 bpm the average man in Tamilnadu lives for about 70 years. Odd man out!

Origin of HRV :

In 1628, William Harvey hinted at a link between the brain and the heart when he wrote: "For every affection of the mind that is attended with either pain or pleasure, hope or fear, is the cause of an agitation whose influence extends to the heart."  For the past half century, numerous anatomic and physiological studies of cardiac autonomic nervous system have investigated this link and found it to be very complex **[Circ. Res. 114 (2014) 1004-1021**].

Heart is not an isolated subsystem. Its dynamics determines and is determined by the rest of the body state. The blood circulation throughout the body, for which the heart is literally the pump, provides a feedback to the heart and the lungs via the autonomous nerves which carry signals from baroreceptors [present in carotid artery] and chemoreceptors [present in carotid body]. Medulla oblongata in the brain stem [that connects the brain to the spinal cord] in the neck region receives and routes the afferant and efferant nerve signals. The influence of these controls on the heart rate, blood pressure and the respiration rate have been modeled by delay differential equations [**J**. *Math.Biol.* **51(2005) 508-526, J**. *Math. Biol.* **36(1997) 41—63**, Rev *Syst Biol Med.* **2012 March; 4(2): 163–170, Science 197(1977) 287-289**].

There is a large body of literature on the connections between the autonomous nervous system [ANS] and the heart and HRV on the one hand and how the ANS brings about a multi-organ coupling and hence a coordination among the heart, the lungs and the blood systems [to list a few].

The Role of ANS in Health and Disease:

Decreased vagal activity after myocardial infarction results in reduced heart-rate variability and increased risk of death. Deceleration capacity is a better predictor of risk than left-ventricular ejection fraction (LVEF) [**Lancet 367(2006)1674–81**]. Cardiovascular reflex tests have shown both sympathetic and para-sympathetic failure in Parkinson's disease [**J Neurol. Neurosurg. Psychiatry 70(2001) 305–310**].

*Many clinical and basic studies [**Journal of Cardiology 59(2012) 117-122**] have demonstrated that abnormal activation of sympathetic nervous system (SNS) is caused by excitory inputs from peripheral baroreceptor and chemoreceptor reflexes and chemical mediators that control sympathetic outflow. The renin-angiotensin system – oxidative stress axis is also in recent focus. Inhibition of the activated SNS using beta-blockers and vagal stimulation have beneficial effects on heart failures. Many has viewed heart failure as an ANS Dysfunction. Direct recordings are available of sympathetic nerve activity using micro-neurography, baroreflex control of blood pressure and blood flow in conditions ranging from rest to postural changes, exercise, and mental stress in populations ranging from healthy controls to patients with hypertension and heart failure. [**Muscle Nerve, 36(2007) 595–614**]. A Good Review. Role of the Autonomic Nervous System in Modulating Cardiac Arrhythmias is studied in [**Circ. Res. 114 (2014) 1004-1021**]. The autonomic tone is related to clinically significant arrhythmias: atrial fibrillation, ventricular fibrillation and ventricular tachyarrhythmias etc. The effect of vagal and sympathetic tones on the cardiac rate was noted as early as 1934 [**Am J Physiol. 110(1934) 42-45**]. Heart Rate Variability has been used as a Predictor of Autonomic Dysfunction in Patients Awaiting Liver Transplantation [**Digestive Diseases and Sciences, 45 (2000) 340 – 344**]. Standard Deviation (in m sec) and ApEn were less in patients with liver disease. PNN50 (%) was distinctly less in patients with liver disease, all compared to control group.*

*In "Monitoring Severity of Multiple Organ Dysfunction Syndrome: New Technologies" [**Pediatr. Crit. Care Med. 2017 March; 18(3 Suppl. 1): S24–S31**], biomarkers are used to assess and monitor the severity and progression of multiple organ dysfunction syndrome (MODS) in children. HRV, Respiratory Rate Variability [RRV], sympathetic drive, oxidative stress [as measured by whole blood redox potential], sepsis and infection and inflammation are some important markers. It may also be argued that, for better outcomes from Ventilator Supports, the Ventilators may have to be tuned to generate the optimal RRV that matches the HRV as RRV and HRV are naturally coupled.*

*<u>Signatures of ANS control in HRV and BP:</u> There are two phenomena, respiratory sinus arrhythmia [RSA] and Mayor Waves, which are widely linked to ANS function [RSA at 0.25Hz and Mayor Waves at 0.1Hz in the power spectrum of the RR-intervals]. RSA is a naturally occurring variation in heart during the breathing cycle [note: Breathing is directly controlled by the brain]. Inhalation reduces the vagal activity causing increase in heart rate. Exhalation causes the vagal activity to*

*resume and hence the heart rate decreases. Thus one has a respiratory sinusoidal component added to vagal firing rate. This affects directly the heart rate and indirectly the BP as BP is coupled to the heart rate. Baroreflex time delay is shown to produce Mayor Waves.*

*[A set of 2 non-linear delay ODEs in time for heart rate and BP modulated by sympathetic and vagus nerves activated by baroreflex have been studied. Respiration is added by hand to the BP variable to produce RSA. Baroreflex time delay is shown to produce Mayor Waves. The time delay and levels of sympathetic and vagal activity could provide useful diagnostic information about the state of the cardiovascular system.* **["Confronting a Cardiovascular System Model with Heart Rate and Blood Pressure Data"] [Article in public domain by PE McSharry, MJ McGuinness and AC Fowler]***]*.

*A very interesting debate titled "Cardiovascular variability is/is not an index of autonomic control of circulation" is to be found in* **Journal of Applied Physiology, 101(2006) 676–682**. *Another interesting read is "Neural influences on cardiovascular variability: possibilities and pitfalls" [***Am J Physiol. Heart Circ. Physiol. 282: H6–H20, 2002***].*

*There is also evidence for a central origin of the low-frequency oscillation in RR-Interval variability" [***Circulation 98(1998) 556-561***]. In this work spectral density using FFT is reported for RR-intervals, SAP (Systolic Arterial Pressure) and Respiration (RRV) of 2 CHF patients before and after fixing LVAD (Left Ventricle Assist Device). No LF at ~.1Hz for Pre-LVAD in RR, SAP and RRV though HF at ~0.25 is present. This absence of LF marks CHF. LF reappears in RR after fixing LVAD, though not in SAP and RRV. Reappearance of LF in RR and not in SAP is taken to imply that LF has more to do with the central autonomic outflow than baroreflex control.*

*Effects of anesthetic, autonomic blockade, apnea and exercise on HRV, Respiration and their coordination are all studied experimentally in detail [***M. M. Kabir, "Detection of Cardiorespiratory Interaction for Clinical Research Applications", Thesis, University of Adelaide, Australia, 2007 ; Circulation. 129( 2014) 2100–2110 ; Circ Res. 19(1966) 400-411***]. Interesting connections exist between heart and lung dynamics [***Journal of Sports Science and Medicine 9(2010)110-118***]. Oxygen uptake increases linearly with Ventilation and Heart Rate also increases linearly with Oxygen uptake. Synchronization in the human cardiorespiratory system [***PHYSICAL REVIEW E 60(1999) 857-870***] is studied using a statistical approach and compared to synchronization in noisy and chaotic systems. Effects of phase and frequency locking between ECG and Respiratory flow are also modeled.*

Now we turn to the model framework.

**The Point-Process-Based Model**

At the outset the model that we consider for the sympathetic and parasympathetic nerves and the SA node is an effective model in the following sense: these nerves are not contacting the heart and the SA node at just one point but at several points and the intrinsic cardiac trigger though originates from the SA node actually results from the synchrony of a group of specialized cardiac cells possessing "automaticity". Therefore our model has 3 effective objects corresponding to sympathetic nerve contacts and parasympathetic nerve contacts to the heart and the SA node. This effective model is justifiable thus: synchronization might help the different branches of sympathetic (or parasympathetic) nervous system to act in unison. Alternatively, even in the absence of synchronization, what the heart will eventually "see" is the summed up effect of all the neuronal spike trains travelling down the different branches of sympathetic (or parasympathetic) nervous system. This makes good sense even within the point-process theory as one can always add several point-processes to obtain one resultant point-process. Thus our effective model consists of 3 objects each endowed with an effective spike-train and we have 3 point-process rates designated as $\lambda_1$, $\lambda_2$ and $\lambda_3$. These rates may be constant in time, varies with time or even controlled by history and feedback loops as we will see shortly.

The basic objective of this paper is to develop a point-process based framework to combine the neural spike trains coming down the effective autonomous nerves to the effective SA node of the heart and the intrinsic activity of the SA node itself and to generate the sequence of heart beats and specifically the time series of the R-waves and thereby the RR-interval sequence. As discussed in the introduction these neural spike trains carry the end-level information which in turn depend on a host of higher levels of neuronal information transfer and processing. As a result the rates or intensities of these spike trains ($\lambda_1$, $\lambda_2$ and $\lambda_3$) which go into the point-process theory may in general be time, memory and history dependent with feed-back loops from the heart itself. We have before us several scenarios:

I. Constant rates independent of time

We said in the Introduction that the internal and external factors determine the values of these neural firing rates. Therefore, for the physiological state of the subject where these factors remain fixed in time, these rates may be taken to be constant in time and stationarity will hold. We may define this state as the resting state. This state may be closely approximated if there is no emotional disturbance, no physical/mental stress or activity and no subconscious disturbances like violent dreams if the subject is asleep. These conditions apply only to healthy subjects. In disease, though the above list of conditions may be satisfied, the internal physiological state of the subject may be constantly disturbed by the underlying disease even when the subject is fast asleep! However one may argue that there are always physiological processes like respiration and other natural rhythms and even ever-present feed-backs may invalidate this constant-rate assumption. We agree. The point of treating this case is to gain some insight and besides it is analytically tractable as shown in the Methods Section. In point-process terminology this model corresponds to the assumption of homogeneous Poisson processes for the input spike trains at the SA node.

II. Time-dependent rates without history and feedbacks

This case may be more realistic compared to Case I. Here the rates are given functions of time. As we know that heart, lungs and circulation are coupled through baroreceptors and chemoreceptors and there are always neural signal delay times, history and feedbacks cannot be set aside completely. Though an analytical framework could be hard to achieve, a Monte Carlo simulation with a priori known time-dependent rates is feasible using the time-rescaling theorem or the thinning algorithm [**Neural Computation 14(2001) 325-346**].

III. Rates with history and feedbacks and time delays

This is the most general and important case which subsumes cases I and II as special cases. As already stated heart dynamics, blood circulation and respiration have feed-back loops mediated by baroreceptors and chemoreceptors and this mediation also involves time delays along afferent and efferent nerves giving rise to history-dependence.

Case I and Case III will be dealt with in detail in this paper.

## Case-I

The model starts with 3 effective point processes (denoted 1, 2, 3), also known as spike trains in the field of neuroscience, that may be pictorially depicted as follows.

```
1 ->    | | |   |  | |  ||| |  ........     Sympathetic Spike Train      [λ₁ and t₁]
2 ->    | | |   |  | |  ||| |  ......       Intrinsic Spike Train        [λ₂ and t₂]
3 ->    | | |   |  | |  | | |  |...         Para-sympathetic Spike Train [λ₃ and t₃]
```

The framework detailed in the Methods Section below arrives at the multivariate probability density for the RR-interval series for the heart beat sequence which may be represented as

```
 |  |    |   |   |   |          |    |   |   | |......
$t_R(1)$              $t_R(k)$      $t_R(k+1)$
```

The multivariate probability density for the RR-interval series has a closed analytic representation: [see the Methods Section for notation and note further that the equation numbering starts in the Methods Section and continues here]

$$P(RR_1, RR_2, \ldots, RR_N) \prod_{k=1}^{N} RR_k = \prod_{k=1}^{N} \exp(-(\lambda_1 + \lambda_2 + \lambda_3)RR_k)$$

$$\prod_{k=1}^{N} \begin{bmatrix} H(RR_k - T_1) \exp((\lambda_1 + \lambda_2 + \lambda_3)T_1)\lambda_1 \\ + H(RR_k - T_2) \exp((\lambda_1 + \lambda_2 + \lambda_3)T_2)\lambda_2 \\ + H(RR_k - T_3) \exp((\lambda_1 + \lambda_2 + \lambda_3)T_3)\lambda_3 \end{bmatrix}$$

$$\prod_{k=1}^{N} RR_k \qquad\qquad [16]$$

For constant lambdas, the random variables $RR_k$ (k=1 to N) constitute a set of independent and identically distributed r.v.s. For time-dependent lambdas, as for inhomogeneous Poisson for example, they are still independent though not identically distributed. However the independence too will be lost for history-dependent processes though independence can still be reclaimed using the time-rescaling theorem and the transformed variables. We may also extract RR interval sequences corresponding to these r.v.s and study their properties using statistical and fractal measures such as multi-scale entropy and Hurst exponent. A typical histogram of an RR interval sequence generated by our MC code was examined and interestingly the peak positions nearly correspond to the values of $T_1$, $T_2$ and $T_3$ used in the simulation.

**Utilities of the Constant Rate Model**

The models advanced in above section, both analytic and Monte Carlo, may be used in several ways. First, one may extract the lambda-s and T-s from any given RR interval patient data using non-linear parametric fitting of the data to the model. Here we have two options: (A) Use the data and the MC algorithm to generate the objective function and (B) Use the data and the analytic model framework to carry out Maximum Likelihood Estimation of the lambdas provided the T-s are known a priori. The values of $T_1$, $T_2$ and $T_3$ themselves may be obtained from the histograms of the RR interval data provided we discern well defined peaks in the histograms. In such a case, even the lambda-s could be extracted from the RR-interval histogram as explained below without resorting to more elaborate data analysis procedures like (A) and (B) mentioned above. As the values of the lambda-s are good indicators of the internal states of the subject in health, disease and in varied kinds of stress, emotions etc., an analysis based on the present model will shed much light on how the internal physiological condition manifest in the RR interval data. The values of $\lambda_1$ and $\lambda_3$ are actually the quantitative markers of the sympathetic and vagal tones respectively.

**Labelled ECG**

As alluded to in the Methods Section and looking ahead into probable future developments in this field one should anticipate that in the very near future we will have ECGs where each PQRST wave is labeled as arising from sympathetic, intrinsic or para-sympathetic activation. In such a scenario, the model predicts the following analytical results for $\lambda_1$, $\lambda_2$ and $\lambda_3$.

$$\lambda_1 = \frac{n_1}{T - n_1 T_1 - n_2 T_2 - n_3 T_3} \quad [17]$$

$$\lambda_2 = \frac{n_2}{T - n_1 T_1 - n_2 T_2 - n_3 T_3} \quad [18]$$

$$\lambda_3 = \frac{n_3}{T - n_1 T_1 - n_2 T_2 - n_3 T_3} \quad [19]$$

Where $n_1 + n_2 + n_3 = N$ is the total number of beats and $T$ the total time of the ECG recording. The above formulas are based on maximum likelihood estimation as well as probabilistic reasoning. For such labeled ECG's equation 16 will further simplify to the following:

$$P(RR_1, RR_2, \ldots, RR_N) = \prod_{k=1}^{N} \exp(-(\lambda_1 + \lambda_2 + \lambda_3)RR_k)$$
$$\times \exp(n_1(\lambda_1 + \lambda_2 + \lambda_3)T_1)$$
$$\times \exp(n_2(\lambda_1 + \lambda_2 + \lambda_3)T_2)$$
$$\times \exp(n_3(\lambda_1 + \lambda_2 + \lambda_3)T_3)$$
$$\times (\lambda_1)^{n_1}(\lambda_2)^{n_2}(\lambda_3)^{n_3} \quad [20]$$

One can show that the probability that any given heart beat has originated from activation of the kind $i$, where $i$ stands for sympathetic when $i = 1$, intrinsic when $i = 2$ and parasympathetic when $i = 3$, is given by

$$P_i = \frac{\lambda_i}{(\lambda_1 + \lambda_2 + \lambda_3)} \quad [21]$$

One can further show that

$$\frac{n_i}{\lambda_i} = \frac{N}{\lambda_1 + \lambda_2 + \lambda_3}$$ which is independent of the index i. [22]

One may make a dissection of equation 16 if we have the ECG of a real patient as follows:

Suppose that we have his ECG record of N heart beats. Each of these heart beats will contribute a factor to equation 16. If this beat has an RR interval, as measured from the previous beat, greater than $T_1$ but less than $T_2$, then this beat will contribute the factor:

$$\exp(-(\lambda_1 + \lambda_2 + \lambda_3)RR)f_1 \quad [23]$$

A beat with an RR interval greater than $T_2$ but less than $T_3$ will contribute the factor:

$$\exp(-(\lambda_1 + \lambda_2 + \lambda_3)RR)(f_1 + f_2) \quad [24]$$

A beat with an RR interval greater than $T_3$ will contribute the factor:

$$\exp(-(\lambda_1 + \lambda_2 + \lambda_3)RR)(f_1 + f_2 + f_3) \quad [25]$$

Pooling all these factors together equation 16 may be simplified to:

$$\exp(-(\lambda_1 + \lambda_2 + \lambda_3)T)(f_1)^n (f_1 + f_2)^l (f_1 + f_2 + f_3)^m \quad [26]$$

Where n, l and m are the number of heartbeats falling in the three categories just mentioned. T is the total time duration of the ECG recorded and

$$f_1 = \exp((\lambda_1 + \lambda_2 + \lambda_3)T_1)\lambda_1 \quad [27]$$

$$f_2 = \exp((\lambda_1 + \lambda_2 + \lambda_3)T_2)\lambda_2 \quad [28]$$

$$f_3 = \exp((\lambda_1 + \lambda_2 + \lambda_3)T_3)\lambda_3 \quad [29]$$

For example the values of n, l and m for a typical ECG [nsr001, PHYSIOBANK ATM] record of length 1000 heartbeats are respectively 589, 409 and 2.

One may further observe: (i) the minimum RR interval for any subject should be greater than $T_1$. Otherwise the above product will vanish as there will then be at least one vanishing factor (ii) if para-sympathetic activation is present in the ECG then the maximum RR interval should be greater than $T_3$. In the data nsr001, RRmin=.531 and RRmax=.852. $T_1$, $T_2$ and $T_3$ values used are respectively .508, .65 and .8. Hence conditions (i) and (ii) are satisfied.

**VARIATION OF HEART RATES WITH THE NERVE FIRING RATES**

We are also able to obtain analytical results for the moments of the RR interval predicted by the present model. The first moment that is the average RR interval can be shown to be:

$$<RR> = \frac{1 + \lambda_1 th_{sy} + \lambda_2 th_{int} + \lambda_3 th_{psy}}{\lambda_1 + \lambda_2 + \lambda_3} \quad [30]$$

Let us now examine how sensitively the heart rate <HR> depends on the 3 nerve firing rates $\lambda_1$, $\lambda_2$ and $\lambda_3$.

$$<HR> = \frac{1}{<RR>} = \frac{\lambda_1 + \lambda_2 + \lambda_3}{1 + \lambda_1 th_{sy} + \lambda_2 th_{int} + \lambda_3 th_{psy}}$$

$$\frac{d<HR>}{d\lambda_1} = \frac{1 + \lambda_2(th_{int} - th_{sy}) + \lambda_3(th_{psy} - th_{sy})}{D^2}$$

$$\frac{d<HR>}{d\lambda_2} = \frac{1 + \lambda_1(th_{sy} - th_{int}) + \lambda_3(th_{psy} - th_{int})}{D^2}$$

$$\frac{d<HR>}{d\lambda_3} = \frac{1 + \lambda_1(th_{sy} - th_{psy}) + \lambda_2(th_{int} - th_{psy})}{D^2}$$

Where $D = 1 + \lambda_1 th_{sy} + \lambda_2 th_{int} + \lambda_3 th_{psy}$

It is interesting to note that the heart rate always increases with the sympathetic activity whereas the heart rate may increase or decrease with the intrinsic activity depending on the strength of the sympathetic and parasympathetic activities. On the other hand the heart rate almost always decreases with the parasympathetic strength. It is also experimentally known that the average RR interval increases with vagal activity [**Journal of the Autonomous Nervous System, 11, 1984, 226-231**]. This also is verified by running the Monte Carlo code and plotting how the average RR interval, which is the reciprocal of the heart rate, varies with one of the nerve activity after keeping fixed the other two nerve activities. Another striking observation is that this variation of the average RR interval with the nerve activity is the same for these two cases: I. The three nerve activities are fed in parallel to the heart and II. They are fed into the heart after pooling them. This is in consonance with the fact that the RR interval histograms for these two cases are identical within numerical errors. One may also show analytically that the above two cases possess the same multi-variate probability density though the Monte Carlo beat generator generates beat sequences with different Hurst exponents!

Another interesting finding deserves a mention here: As you may recall 3 point processes are fed into the SA node of the heart in our model. I have considered two variants of this model: I. the 3 point-processes are pooled together into one single point process which is then fed into the SA node and II. In this variant we do not pool them together but feed them individually into the SA node. Thus we have designed two different generators for the heart-beat sequence. For the same set of values for the parameters in the two beat generators we end up with two different Hurst fractal measures for the resulting beat-sequences [.5799 FOR POOLED AND .4682 FOR DEPOOLED]. The interesting finding is that the RR-interval histogram is identical for the two sequences! See the 4 figures below. A kind of degeneracy!! In the case of energy level degeneracy we know that two or more states sharing the same energy are connected by a symmetry (or symmetries) of the Hamiltonian which generates the degenerate states. Extending this principle to our model we need to ask: What kind of symmetries in the two beat generators lead to identical histograms but two distinct fractal signatures?

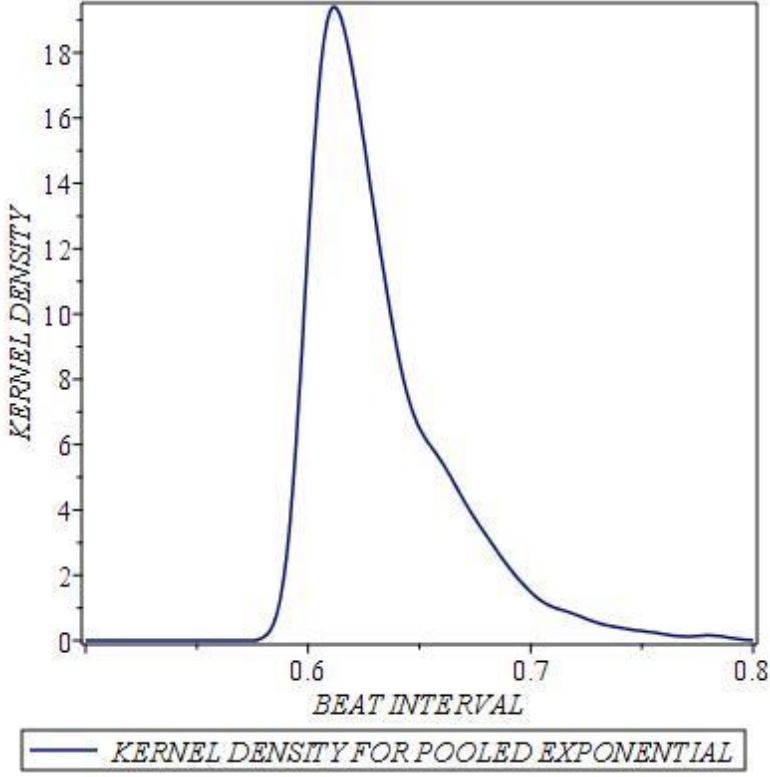

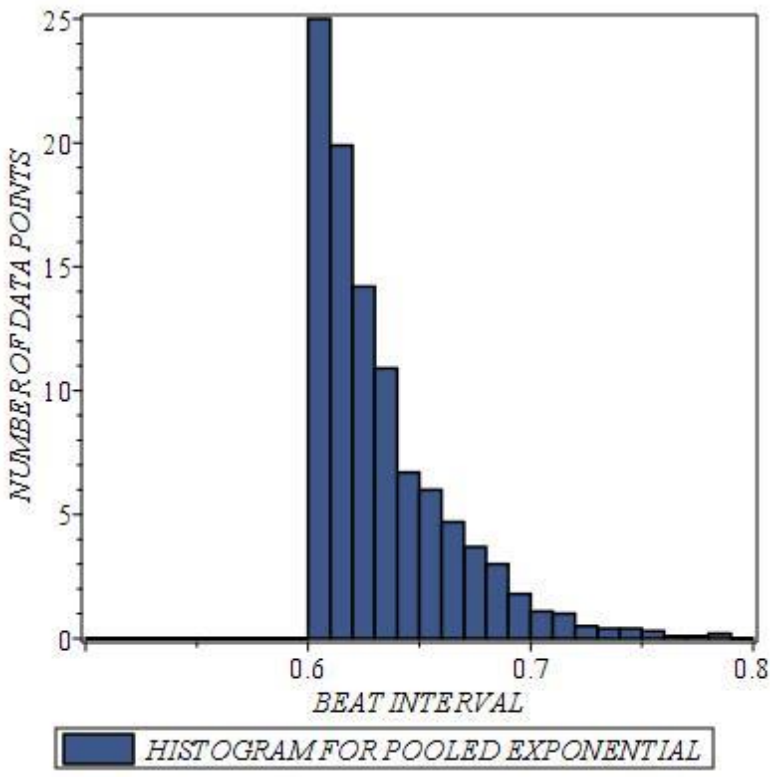

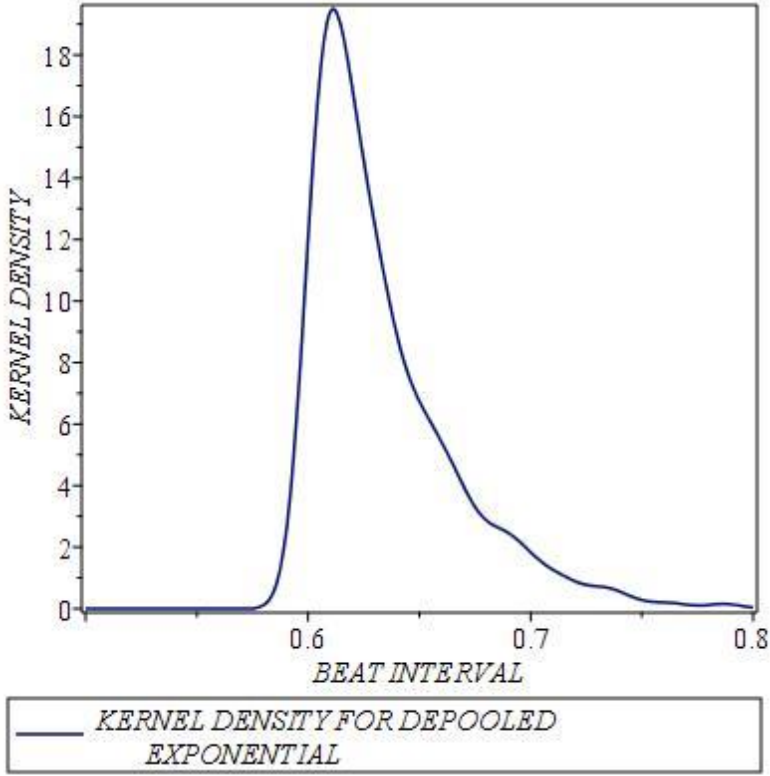
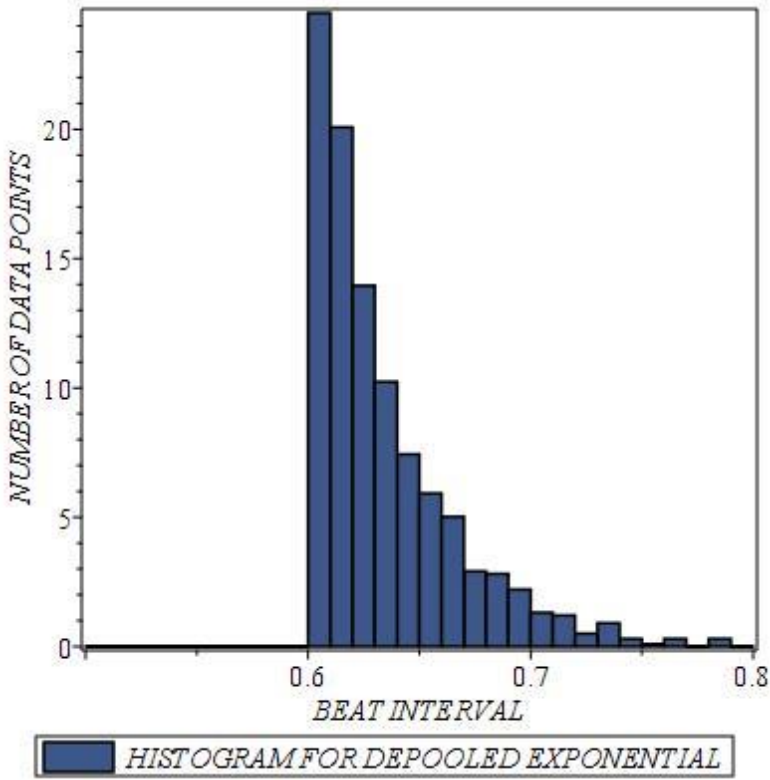

Below we list the Hurst exponents for six different cases for pooled and de-pooled exponential waiting times with fixed threshold times and with threshold times having inverse Gaussian and Pareto distributions:

**HURST EXPONENTS FOR POOLED AND DE-POOLED PROCESSES WITH FIXED AND INVERSE GAUSSIAN AND PARETO DISTRIBUTED THRESHOLD TIMES**

| | |
|---|---|
| POOLED EXPONENTIAL WITH FIXED THRESHOLD TIMES | .5799 |
| DE-POOLED EXPONENTIAL WITH FIXED THRESHOLD TIMES | .4682 |
| POOLED EXPONENTIAL WITH THRESHOLD TIMES DISTRIBUTED AS INVERSE GAUSSIAN | .4492 |
| DE-POOLED EXPONENTIAL WITH THRESHOLD TIMES DISTRIBUTED AS INVERSE GAUSSIAN | .4985 |
| POOLED EXPONENTIAL WITH THRESHOLD TIMES DISTRIBUTED AS PARETO | .5609 |
| DE-POOLED EXPONENTIAL WITH THRESHOLD TIMES DISTRIBUTED AS PARETO | .4684 |

Of note is the fact that pooled exponential waiting time with fixed threshold times and with Pareto-distributed thresholds have Hurst exponents above .5 while for all others it is below .5. However it is near unity for none of the cases. Therefore one may conclude that if we insist on a Hurst exponent near 1 for healthy heart function we need to fine-tune the framework by including mechanisms with history and feedback in the heart dynamics. This we do below.

**MODEL WITH FEEDBACK**

We now propose to include in our model the History (equivalently the Memory) and feed-back loops coming from the baroreceptors and chemoreceptors. With this we hope to have a complete point-process model for the heart. In principle this model should be predicting both healthy and a wide spectrum of unhealthy cardiac function. The following is a schematic of this feedback:

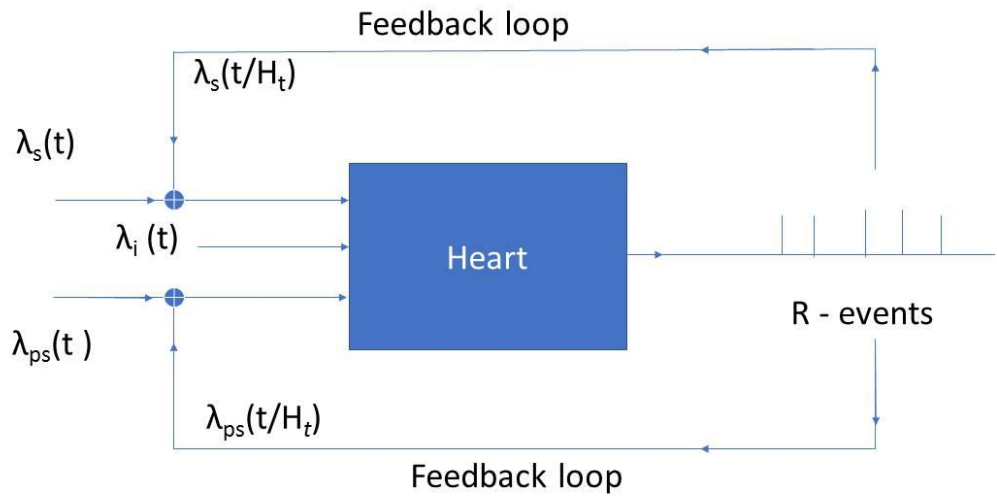

With feedbacks come the most general time-dependence. There are two general methods for handling general time-dependent point processes: Time Rescaling and Thinning. Thinning is recommended for non-homogenous Poisson processes while Time Rescaling is the method of choice for processes with memory and/or feedback. We have used Time Rescaling [**Neural Computation 14(2001) 325-346**].

*BEFORE WE EMBARK ON THAT WE NEED TO KNOW SOMETHING ABOUT THE BARORECEPTORS AND CHEMORECEPTORS AND HOW THEY REGULATE THE HEART DYNAMICS VIA THE AUTONOMOUS NERVOUS SYSTEM. THESE HAVE BEEN STUDIED AND THEIR EFFECTS ON BLOOD PRESSURE, HEART RATES AND BREATHING RATES MODELED USING DELAY DIFFERENTIAL EQUATIONS [DDEs]. THOUGH THESE DDEs ARE VERY INTERESTING IN THEMSELVES AND WE DO HAVE GENERATED SOME RESULTS USING THEM [THOUGH I WILL NOT BE PRESENTING THEM HERE], THEY ARE A KIND OF CONTINUUM DESCRIPTION WHEREAS WHAT WE WANT TO UNDERSTAND IS DISCRETE HEART BEATS.*
*THE BARORECEPTORS PRESENT IN THE CAROTID ARTERY SENSE THE BLOOD PRESSURE WHICH IN TURN IS CONTROLLED BY THE HEART. AFFERENT NERVES CARRY THE PRESSURE SIGNALS TO THE BRAIN STEM AND EFFERENT NERVES [SYMPATHETIC AND VAGUS] INFORM THE HEART ABOUT THE BLOOD PRESSURE VARIATION. THIS HELPS IN MAINTENANCE OF OPTIMAL HEART RATE AND HENCE BLOOD PRESSURE.*
*THE CHEMORECEPTORS SENSE THE CO2 LEVEL IN THE BLOOD AND PASS IT ON TO THE ANS THAT REGULATES THE BREATHING RATE. THE BREATHING RATE IS TIGHTLY COUPLED WITH THE HEART RATE WHICH GO UP OR DOWN TOGETHER.*
*TO HAVE A COMPLETE MODEL WE NEED BLOOD PRESSURE DATA AND CO2 DATA. WE THEN USE THESE DATA TO ARRIVE AT THE FIRING PATTERN IN THE SYMPATHETIC AND PARA-SYMPATHETIC NERVES WHICH FEED THE SA NODE OF THE HEART. IN THE ABSENCE OF SUCH DATA WE MAY PRETEND THAT WE KNOW HOW THE HEART RATE AFFECTS THE BLOOD PRESSURE WHICH IN TURN CONTROLS THE FIRING PATTERN IN THE ANS. HOWEVER, ONCE WE HAVE THE PRESSURE DATA AVAILABLE WE CAN DIRECTLY USE IT WITHIN THE PRESENT FRAMEWORK.*

*FOR THE IMPLEMENTATION OF THE FEEDBACK MODEL ONE NEEDS TO KEEP IN MIND:*
1. *THAT NERVE SIGNAL TRANSMISSION IS NOT INSTANTANEOUS BUT INVOLVES DELAYS WHICH SHOULD BE INCORPORATED IN THE MODEL[DENOTED TAUS AND TAUP IN THIS PAPER]*
2. *BECAUSE OF THE DELAYS IN THE NERVES THE HEART RATES IN THE PAST INFLUENCE THE PRESENT FIRING RATES AT THE SYNAPSE BETWEEN THE NERVE TERMINALS AND THE SA NODE. THIS HISTORY IS TO BE KEPT TRACK OF PROPERLY*
3. *THE NERVE FIRING RATES HAVE TO BE PROPERLY LINKED TO THE HEART RATE USING EMPIRICALLY KNOWN FUNCTIONS*
4. *INCORPORATION OF THESE ELEMENTS IN THE BASIC POINT-PROCESS FRAMEWORK THAT WE ENUNCIATED RESULTS IN THE COMPLETE FEEDBACK MODEL.*

*THE SYMPATHETIC TONE AND THE PARASYMPATHETIC TONE ARE CONNECTED TO THE ARTERIAL PRESSURE BY THESE SIGMOIDAL FUNCTIONS:*

$$T_S(P_a) = \frac{1}{1+\left(\frac{P_a}{\alpha_S}\right)^{\beta_S}} \qquad T_P(P_a) = \frac{1}{1+\left(\frac{P_a}{\alpha_P}\right)^{-\beta_P}}$$

A SIMPLISTIC MODEL WILL CONNECT THE HEART RATES TO THESE TONES THUS: [WE WILL NOT BE EMPLOYING THIS]

$$\dot{H} = \alpha_H T_S(P_a) - \beta_H T_P(P_a)$$

WHERE $\alpha_S = \alpha_P = 93$ mmHg $\qquad \beta_S = \beta_P = 7 \qquad \alpha_H = .84 \quad \beta_H = 1.17$

THE VALUES OF THESE PARAMETERS ARE TAKEN FROM THE LITERATURE [J.Math.Biol. 36(1997) 41-63].

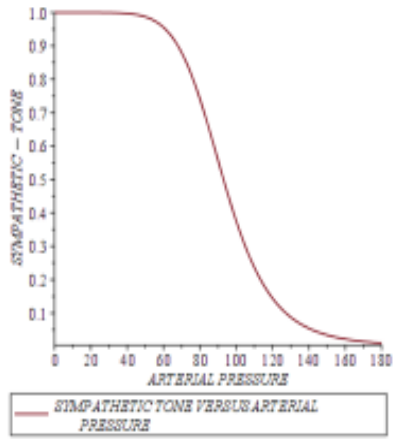
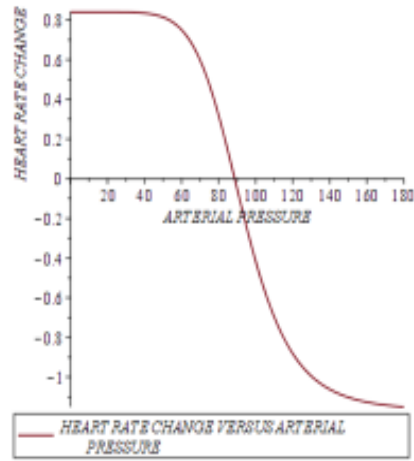
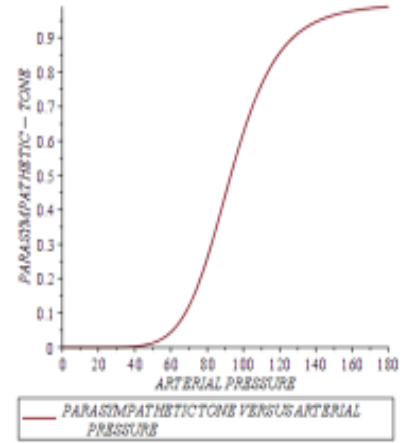

AS REMARKED EARLIER ONE MAY PROCEED DIRECTLY WITH THE ABOVE EQUATIONS IF WE HAVE PRESSURE DATA. IN THE ABSENCE OF PRESSURE DATA ONE MAY APPROPRIATELY RELATE THE PRESSURE DATA TO THE RR INTERVAL DATA AND REWRITE THE ABOVE SIGMOIDAL FUNCTIONS IN TERMS OF RR INTERVALS:

$$T_S(RR) = \frac{1}{1 + \left(\frac{RR}{RRrefS}\right)^{-\beta_S}}$$

$$T_P(RR) = \frac{1}{1 + \left(\frac{RR}{RRrefP}\right)^{\beta_P}}$$

$$RRrefS = \frac{\alpha_S}{\gamma}$$

$$RRrefP = \frac{\alpha_P}{\gamma}$$

$$\gamma = 0.16$$

IT SHOULD BE NOTED THAT THESE TWO SIGMOIDS TAKE NON-DIMENSIONAL VALUES WHICH VARY BETWEEN ZERO AND UNITY AND CAPTURES THE MODULATION OF THE FIRING RATES BY BARORECEPTORS AND CHEMORECEPTORS. TO CONVERT THESE INTO ACTUAL FIRING RATES IN SPIKES PER SECOND WE NEED TO MULTIPLY THESE SIGMOIDS BY WHAT WE WOULD CALL MAXIMAL FIRING RATES OR SIMPLY FIRING RATES OF THE SYMPATHETIC AND PARASYMPATHETIC NERVES. IT IS THESE FIRING RATES THAT WE WILL TREAT AS VARIABLE PARAMETERS IN THE MPDEL AND STUDY THE HURST EXPONENTS AS A FUNCTION OF THESE FIRING RATES HENCEFORTH DENOTED AS SYM AND PARASYM.

THE TWO FIGURES BELOW REPRESENT THESE SIGMOIDS.

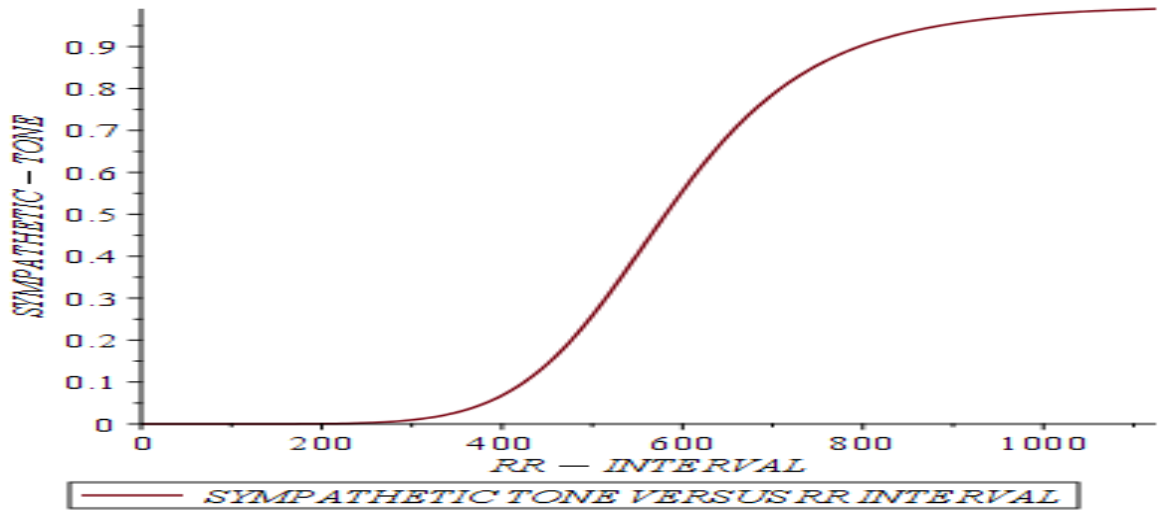

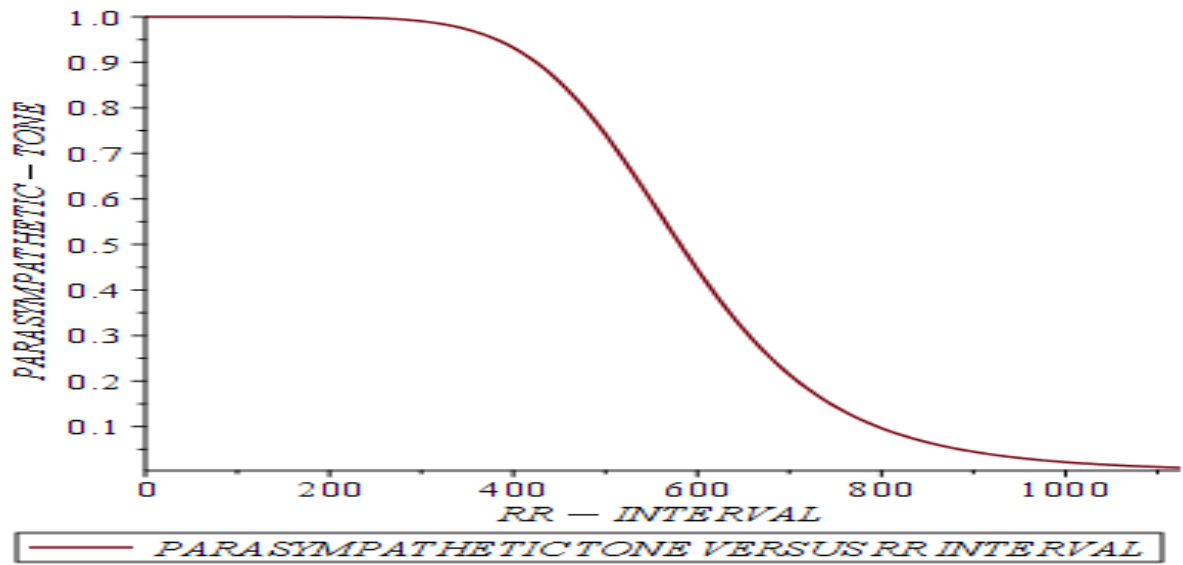

Using the same point-process model that we outlined in the beginning the feedback loops were properly implemented, using the delayed ANS signals which in the general case come from the baroreceptors and chemoreceptors, and heart beat sequences were generated and their Hurst exponents computed for different parameter settings. The results are presented below:

## HURST EXPONENTS FOR A RANGE OF NERVE FIRING RATES

[TAUP=2, TAUS=8]

PARASYM=1

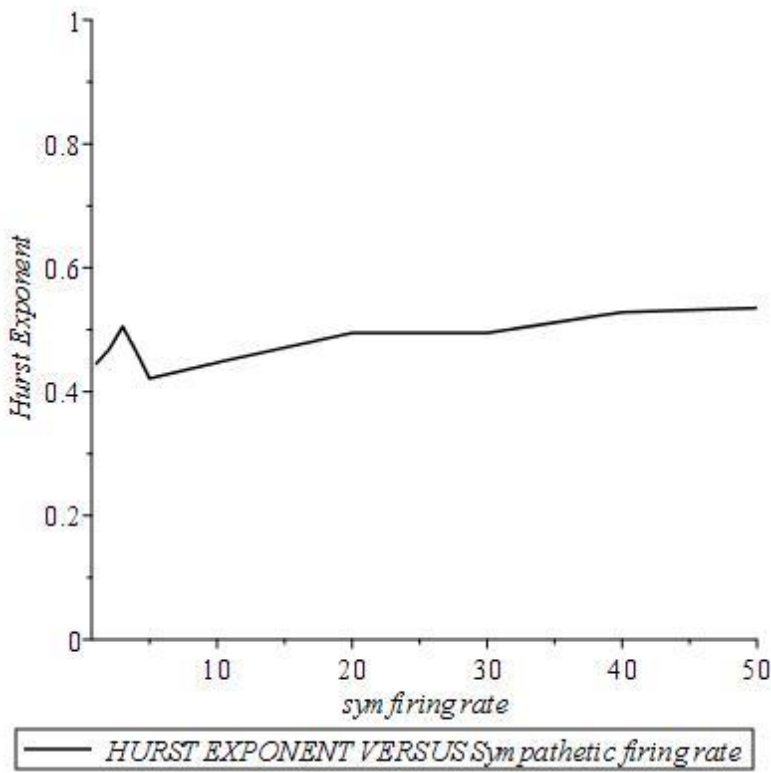

It is clear from this figure that the Hurst exponent hovers around 0.5 when the parasympathetic tone is small at unity and the sympathetic tone increases up to 50. However note the slight increase followed by a tendency to saturate.

PARASYM=30

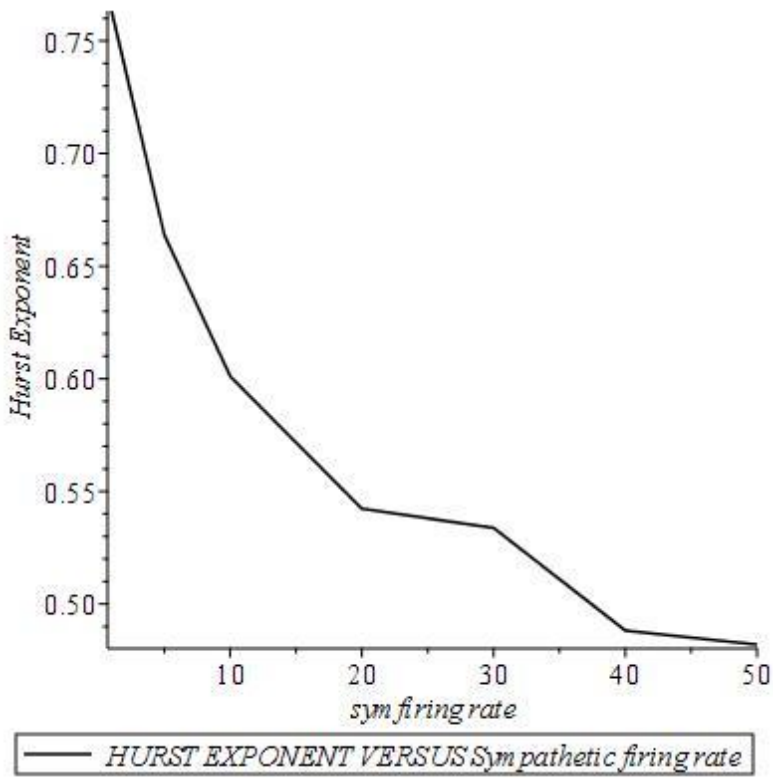

Here the parasympathetic tone is high at 30 and the sympathetic firing rate increases from zero to 50. The Hurst exponent falls rapidly to about 0.5 from an initial high value.

SYM=1

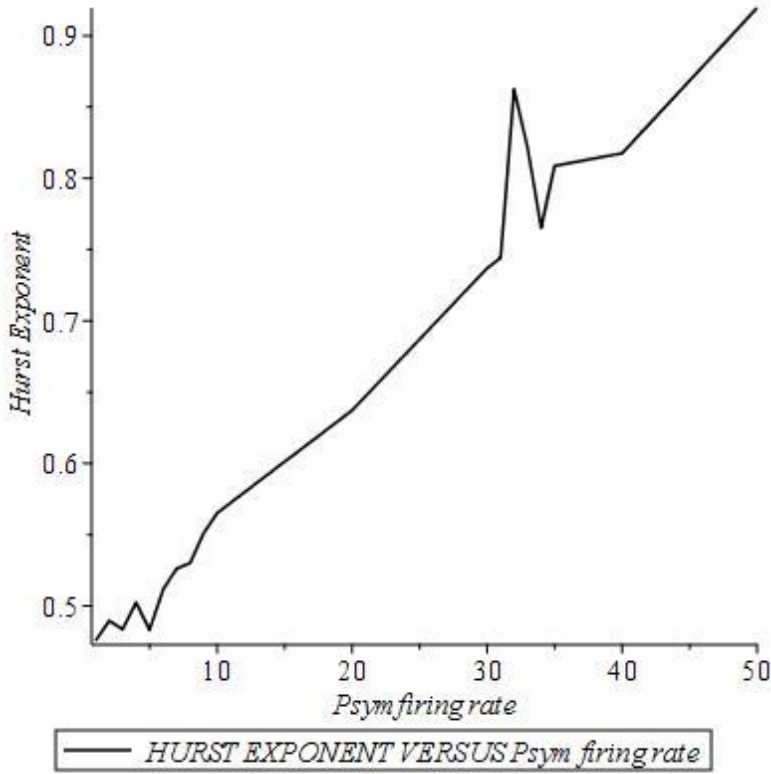

The sympathetic tone is set at unity and the parasympathetic tone is increased from zero to 50. The Hurst exponent increases almost linearly from about 0.5 to near unity. In all such figures we notice some degree of fluctuations though the trends are pretty clear. One reason could be that we are doing Monte Carlo simulations using random number generators and if we repeat these simulations using several random number seeds and average the results the fluctuations should in principle disappear.

SYM=30

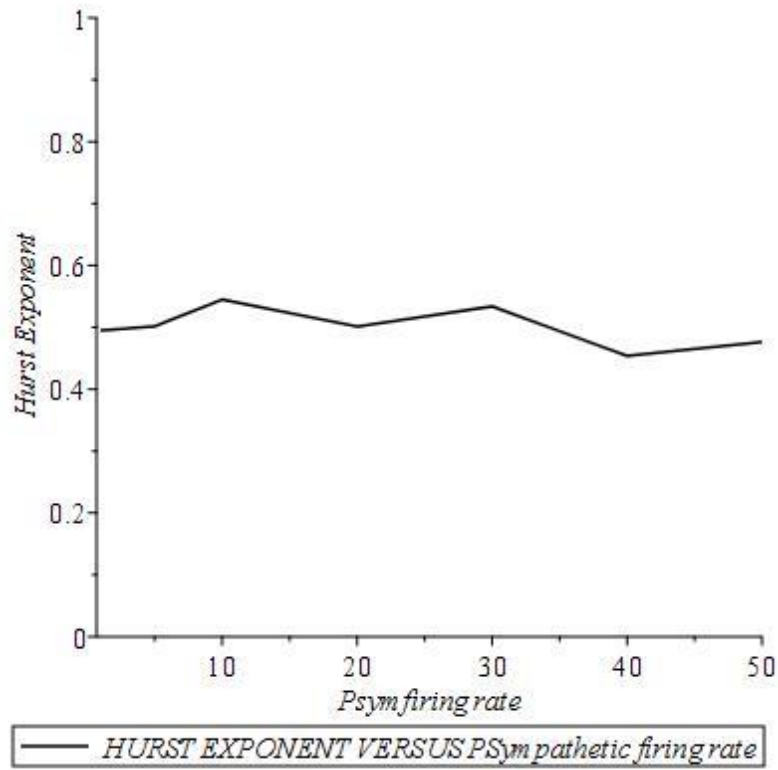

This plot is quite interesting in that the Hurst value is near 0.5 even for high parasympathetic firing rates due to the fact that the sympathetic tone is set high at 30. Clearly this behavior is in sharp contrast to the previous figure where the Hurst value increased with the parasympathetic tone. The moral of the story is that the sympathetic tone tries to bring down the Hurst exponent while the parasympathetic tone tries to boost it up.

# HURST EXPONENTS FOR A RANGE OF NERVE FIRING RATES FOR EQUAL NERVE CONDUCTION TIME DELAYS

[TAUP=6, TAUS=6]

PARASYM=1

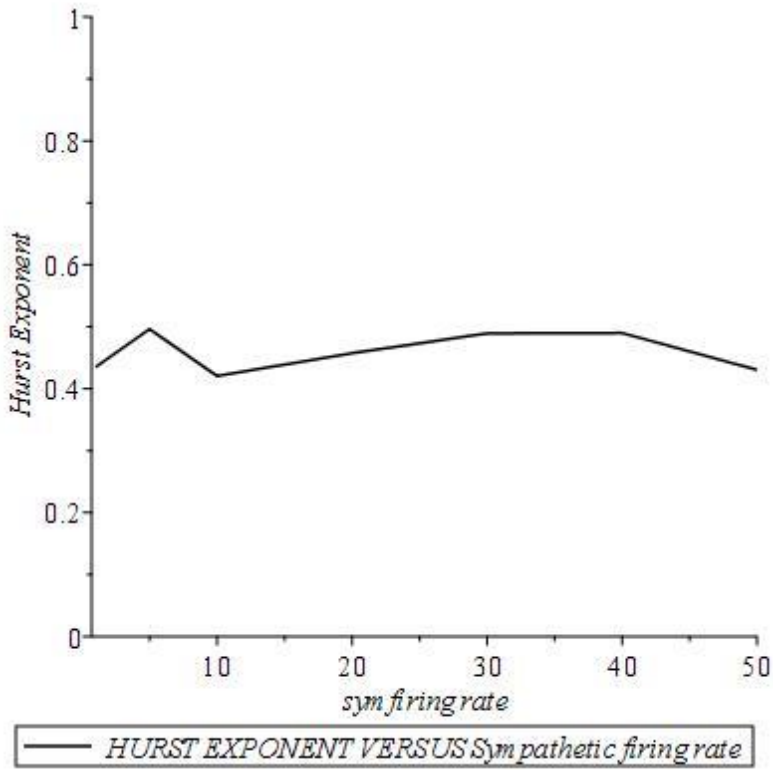

PARASYM=30

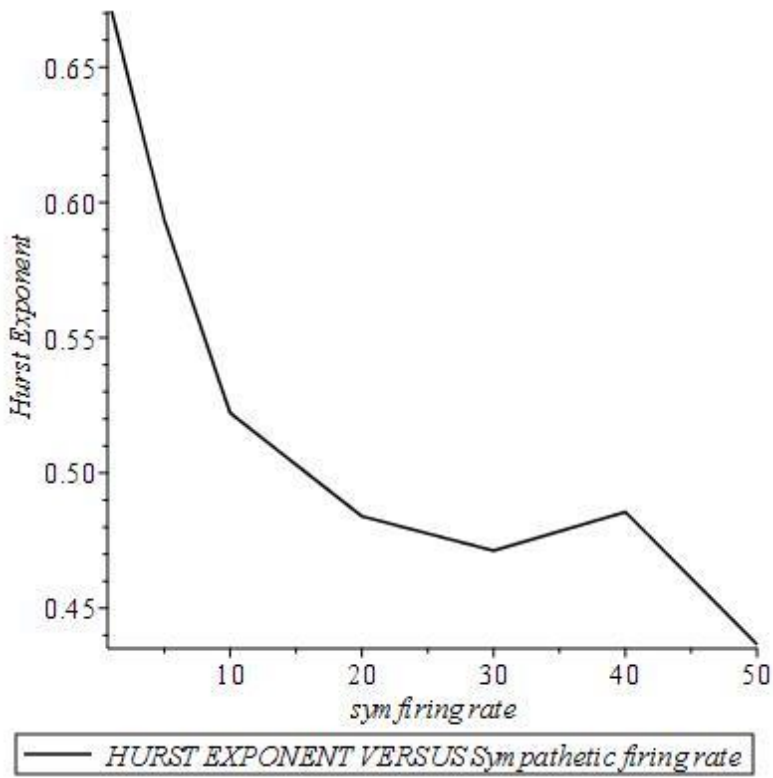

SYM=1

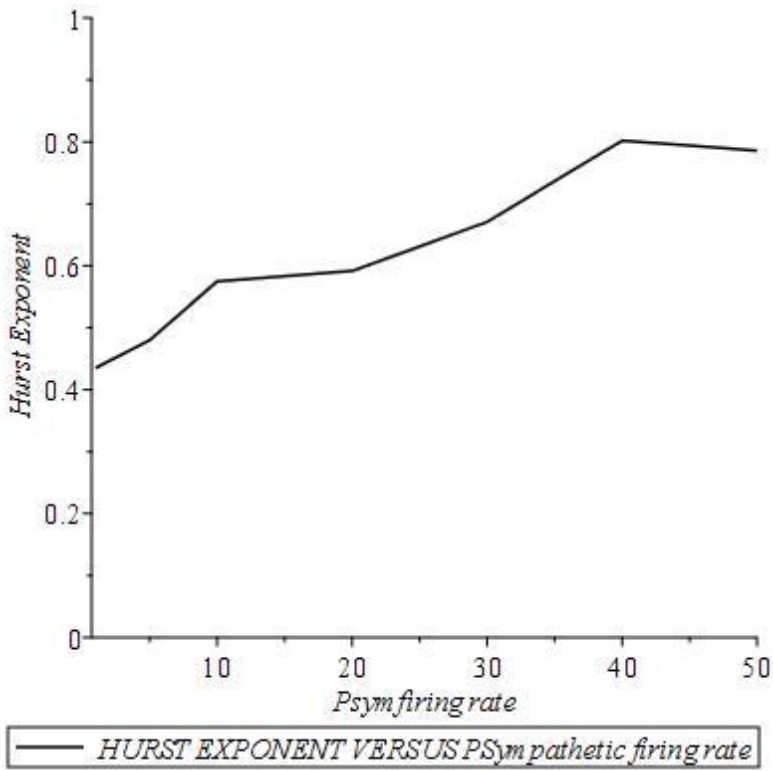

SYM=30

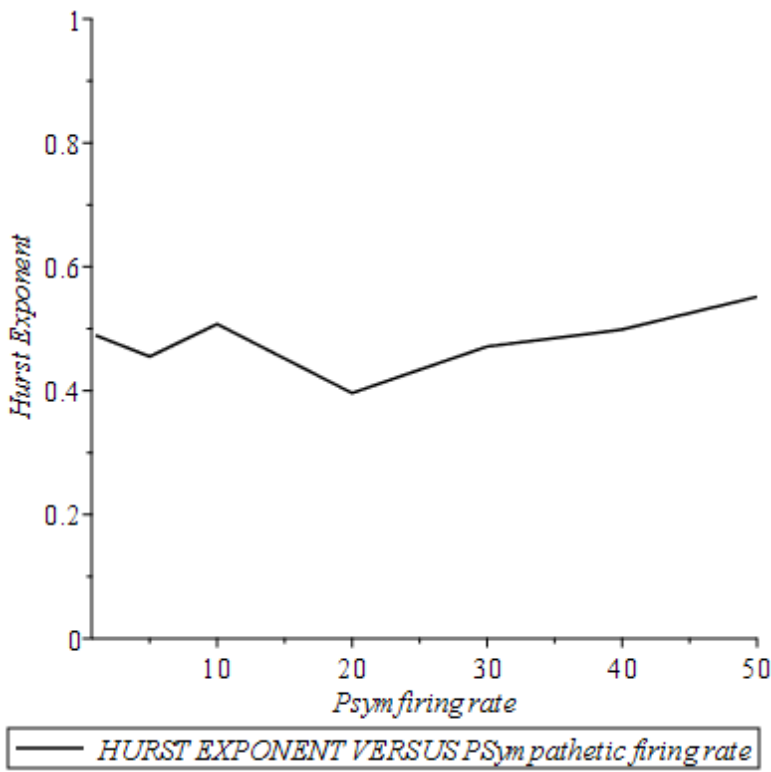

Qualitatively the trends are the same as that of the case with different time delays for the sympathetic and parasympathetic activations of the heart. However the variation of the Hurst exponent with the nerve firing rates are somewhat weakened when the sympathetic and parasympathetic time delays are set equal.

**Denervated Vagus and Sympathetic Nerves**

In animal experiments it has been common to denervate vagus or sympathetic nerves to study the impact on the animal heart function. In human beings denervation by medication has been studied though cutting the nerves may pose its own risks. In simulations such denervations are possible just by zeroing the vagus or sympathetic firing rates. Below Hurst exponent plots are presented after denervating one of these autonomic nerves.

DENERVATED VAGUS

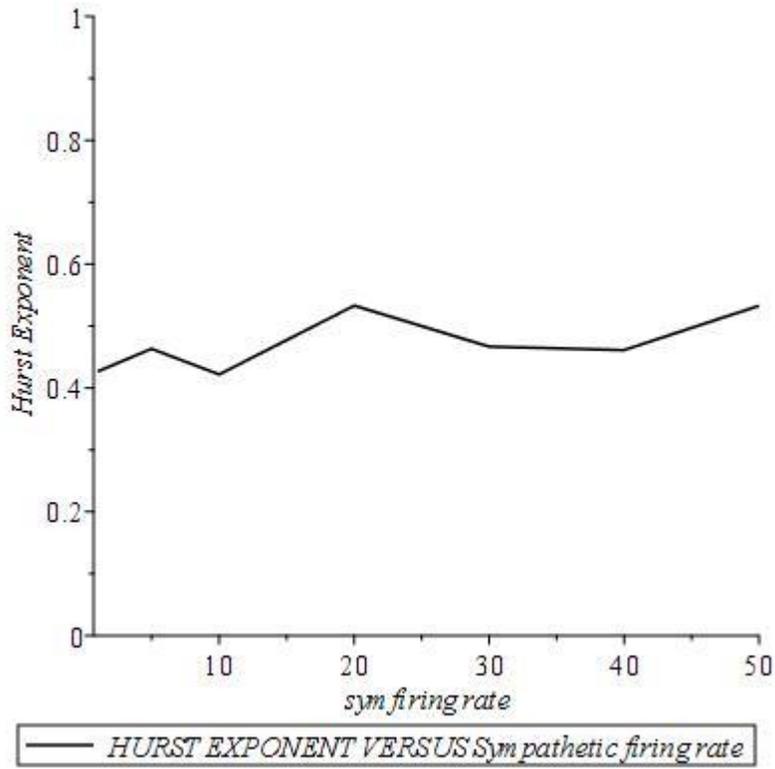

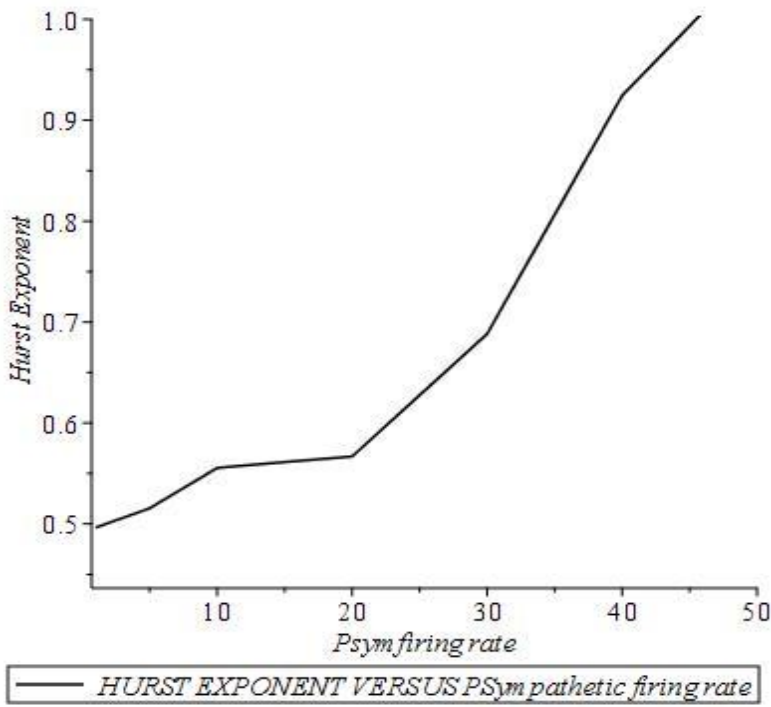

DENERVATED SYMPATHETIC NERVE

It is clear from these plots that the Hurst exponent hovers around 0.5 for the denervated vagus whereas the Hurst exponent rises from about 0.5 to unity when the vagus nerve is active and the sympathetic nerve is denervated. On the other hand, if both the vagus and sympathetic nerves are cut, the Hurst exponent is 0.4996 which is almost 0.5.

## Conclusions and Future Perspectives

We have advanced a point-process based framework for the regulation of heart beats by the autonomous nervous system and analyzed the model with and without feedback. The model without feedback was found amenable to several analytical results that help develop an intuition about the way the heart interacts with the nervous system. However, in reality, feedback, baroreflex and chemoreflex controls are important to model healthy and unhealthy scenarios for the heart. Based on the Hurst exponent as an index of health of the heart we show how the state of the nervous system may tune it in health and disease.

The influence of baroreceptors and chemoreceptors on the heart dynamics is well studied experimentally and their average effect on the heart rate modeled using delay differential equations using continuum approximations. However such descriptions fails to generate a discrete sequence of heart beats which alone provides the basis for any clinical application of the theory for diagnosis. For example fractal scaling, which provides an important marker for the health status of the heart, is based on the discrete RR-interval series. Besides, as discussed in the section on "Prior Research", almost every other marker of cardiac health has as the input only the RR-interval series. Then there are several ad hoc models using non-linear oscillators and others custom-made to generate a sequence of heart beats with PQRST structure and inter-beat intervals all prescribed in advance. Clearly these ad hoc models are severely handicapped when it comes to the understanding of any feedbacks, baroreflex and chemoreflex controls and any mediation by the autonomous nervous system which any real heart exhibits.

The major conclusions of this paper may be stated as follows:

A very general framework is advanced that includes feedback loops, baroreflex and chemoreflex controls and a host of other controls, arising from sensory and the states of emotion, on the neuronal firing rates in the sympathetic and parasympathetic nerves.

We find using this framework that a heart that is not properly coupled and regulated by the ANS does not correspond to a healthy heart in terms of the Hurst signature. A heart that is properly connected to the ANS exhibits a range of Hurst signatures [from 0.5 to 1] depending on the sympathetic and parasympathetic nerve excitations. The trends that we find prove correct the common folklore that excess sympathetic excitations are bad for the heart and the parasympathetic excitations are good for the heart. Using this model one can further theoretically denervate the heart and study the denervated heart dynamics.

Using non-linear parametric fitting routines it is further possible to extract all the parameters which go into the feedback model using real RR-interval series. This way one may relate the ECG to the underlying innervations present in the ANS and other cardiac parameters.

The analytic framework provided here for the simpler model without feedback gives several exact results which are noteworthy and will be applicable in limited contexts where the feedback may not be important. A simple numerical scheme is also proposed to extract the parameters of this simpler heart model from the histogram obtained using any real ECG. The scheme which we have proposed for the labelled ECG makes the analysis even easier because of the analytical results possible for the labelled ECG.

The present work holds much promise for the future as well. In as much as the central and autonomous nervous systems signal and regulate using a complex neuronal network almost every organ and physiological process in the human body one can never overstate the importance of information transmission and processing which continually takes place using the spike-trains that crisscross this neural network. In mathematical terms this complex set of bio-phenomena translates into a theoretical framework of how this rich set of point-processes talk among themselves, combine among themselves and synthesize new point-processes depending upon the target organs which they ultimately control and serve. Though the present work has advanced such a framework specifically for the heart and how it is controlled by the two branches of the autonomous nervous system, it can be easily modified and generalized to study the neural control of any other organ. Of course this modification or generalization will require as inputs physiological information pertinent to the organ concerned and the nerves controlling it.

Keeping in mind the fact that today's ideas become tomorrow's reality, we speculate on how methods such as the one developed in this paper may help extend in the near future a helping hand to the poor souls fighting with Parkinson's [**J Neurol. Neurosurg. Psychiatry 70(2001) 305–310**] and Alzheimer's and also to the brain-dead. In all these patients the nervous system-central and peripheral- is deranged in one way or another. Several kinds of derangement are possible: (I) Poor or faulty reception of information from sources external or internal to the body (II) Defective or delayed neural processing of the information received and (III) Compromised transmission of the processed information to the target organs. The provocative question that we would like to ask is that "when the natural nervous system is thus compromised, will it be possible to develop and employ artificial neural stimulation capable of supporting minimum body function?" These artificial neural stimulation should consist in 'artificial spike-trains' with the correct information content for the target organ. Besides we will have to allow for the correct cross-talk among these artificial spike-trains as well as the ability to receive external sensory information. Given the present advances in such fields as neuroscience, deep learning and artificial intelligence, the day is not far away when our speculations will become reality. Prosthetic limbs and artificial cardiac pace-makers which are already prevalent in clinical practice should be considered as two encouraging fore-runners for such anticipated innovations.

**Methods Section**

The model starts with 3 point processes (denoted 1, 2, 3), also known as spike trains in the field of neuroscience, that may be pictorially depicted as follows.

```
1 ->    | | |   |  | |  | | |  |  ........     Sympathetic Spike Train      [λ₁ and t₁]
2 ->    | | |   |  | |  | | |  | ......         Intrinsic Spike Train        [λ₂ and t₂]
3 ->    | | |   |  |  |  | | |   | ...          Para-sympathetic Spike Train [λ₃ and t₃]
```

Each spike train is characterized by a set of two parameters: λ and *t* indicated above.

For a clear exposition of the basic model we assume that these spike trains correspond to three homogeneous Poisson processes, with parameters $\lambda_1, \lambda_2$ and $\lambda_3$, though in principle they can be non-homogeneous Poisson or even more general history-dependent processes [**An Introduction to the Theory of Point Processes, Volume I: Elementary Theory and Methods, D.J. Daley and D. Vere-Jone, Springer, Newyork, 2003**].

The additional parameters $t_1, t_2$ and $t_3$ will shortly be introduced. $\lambda_1, \lambda_2$ and $\lambda_3$ are actually the rates of arrival of the spikes [unit: count per second] known as the firing rates of neurons in neuroscience literature. We further know that the inter-arrival time, also known as the waiting time, is a random variable wt having the exponential distribution for the Poisson process:

$P_1(wt)d(wt) = \lambda_1 \exp(-\lambda_1 wt)d(wt)$ [1]

$P_2(wt)d(wt) = \lambda_2 \exp(-\lambda_2 wt)d(wt)$ [2]

$P_3(wt)d(wt) = \lambda_3 \exp(-\lambda_3 wt)d(wt)$ [3]

As these distributions will be the building blocks of our model, we list them explicitly. In Cardiology, $\lambda_1, \lambda_2$ and $\lambda_3$ may be referred to as sympathetic, intrinsic and parasympathetic tones. In the field of Neuroscience these $\lambda$-s are known as the firing rates of the corresponding neurons.

We shall now explain the parameters $t_1$, $t_2$ and $t_3$. The SA node is the primary pace-maker of the heart. A heart beat is initiated when the SA node fires. This trigger happens when a voltage threshold is crossed by the SA node [IEEE Transactions on Biomedical Engineering, 47(2000) No. 9]

]. Consider now the following 3 scenarios:

(A) A spike arrives at the synapse between the sympathetic nerve terminal and the SA node. The neurotransmitter Norepinephrine (NE) raises the membrane voltage and enables it to cross the threshold. $t_1$ is the time taken for this voltage threshold crossing.

(B) In the absence of any spike arriving at the synapse with the SA node, the intrinsic electrical activity at the SA node, which is also known as automaticity of the SA node, may take charge and boost the voltage past the threshold. $t_2$ is the time for this kind of threshold crossing.

(C) A spike arrives at the synapse between the parasympathetic nerve terminal and the SA node. The neurotransmitter Acetylcholine (Ach) which is released into the synapse pushes the voltage beyond the threshold. $t_3$ is the time taken.

It is known experimentally that $t_3 > t_2 > t_1$. This is to say that the sympathetic activation of the SA node is the fastest, the parasympathetic activation the slowest and the intrinsic activation lies in between the two. We shall call these the threshold times.

After introducing the basic concepts and terminology which enter our model, we now start formulating the model and its objectives. The objective is to advance a model, based on the point-process theory for the ECG time series using, as inputs, the 3 point processes we just described. To be more specific we propose to model in probabilistic or Monte Carlo terms the time sequence of the R-waves in the ECG. The internal structure of the PQRST wave form is not our consideration. However we should keep in mind the possibility that the properties of the PQRST, such as the R-wave amplitude for example, may depend on and hence be indicative of the source of activation of any given heart beat which may be sympathetic or parasympathetic or intrinsic in character. We would have an opportunity to say more on this aspect later in this paper.

Represent the time sequence of the locations of the R-waves in an ECG thus:

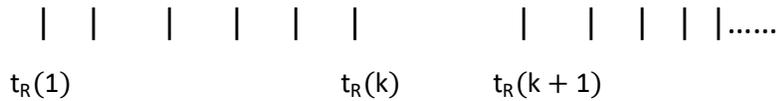

$t_R(1) \qquad\qquad t_R(k) \quad t_R(k+1)$

Let us now focus on the time elapsed between the k-th R-wave at time $t_R(k)$ and the R-wave at $t_R(k+1)$. After the k-th heart beat at $t_R(k)$ the SA node is waiting to be activated by any of the 3 point processes described earlier. If we represent by $wt_k$ the random waiting time just after the k-th heartbeat, $wt_k$ may follow any one of the 3 exponential distributions we wrote down for the 3 kinds of SA node activations. However the only condition is that if the SA node is to be activated, say by the sympathetic nerve, between $wt_k$ and $wt_k + dwt_k$, then clearly the activations coming from intrinsic and parasympathetic routes should NOT happen before $wt_k$. After enforcing this condition, the probability that the activation coming from the sympathetic nerve wins is given by

$$Q_1(wt_k)dwt_k = \exp(-\lambda_2 wt_k)\exp(-\lambda_3 wt_k)\exp(-\lambda_1 wt_k)\lambda_1 dwt_k \quad [4]$$

It is to be noted that the first two factors on the RHS of the above equation gives the probability that the sympathetic and parasympathetic activations do not arrive before $wt_k$. Remember that $wt_k$ is measured from the k-th heart beat time $t_R(k)$. The remaining factors on the RHS give the probability that the sympathetic activation arrives in the time interval between $wt_k$ and $wt_k + dwt_k$. The rule used: probabilities of independent events multiply to yield the joint probability. We believe that this assumption of independence of the 3 point processes is realistic as we have no physiological evidence to the contrary. This assumption will continue to hold even in the case when these point processes are no longer Poissonian but more general history-dependent processes. One can further argue that even in the presence of feed-back loops involving the heart, circulation system and the lungs via the baroreceptors and the chemoreceptors the said assumption holds among the conditional processes in lieu of the original processes.

Following the above procedure, we can write down the corresponding probabilities for the intrinsic and parasympathetic activations respectively:

$$Q_2(wt_k)dwt_k = \exp(-\lambda_1 wt_k)\exp(-\lambda_3 wt_k)\exp(-\lambda_2 wt_k)\lambda_2 dwt_k \quad [5]$$

$$Q_3(wt_k)dwt_k = \exp(-\lambda_1 wt_k)\exp(-\lambda_2 wt_k)\exp(-\lambda_3 wt_k)\lambda_3 dwt_k \quad [6]$$

Here we need to pause for a while in order to connect $t_R(k+1)$ to $t_R(k)$. The connection is this:

For sympathetic activation:

$$t_R(k+1) = t_R(k) + wt_k + t_1 + r_1 \quad [7]$$

For intrinsic activation:

$$t_R(k+1) = t_R(k) + wt_k + t_2 + r_2 \quad [8]$$

For para-sympathetic activation:

$$t_R(k+1) = t_R(k) + wt_k + t_3 + r_3 \quad [9]$$

where $t_1$, $t_2$ and $t_3$ are the thresholding times defined earlier. Without loss of generality, we may set $r_1 = r_2 = r_3 = r$ where r is the cardiac refractory period which is the time period during which a heart which has already initiated a beat cannot entertain another activation. We will further use the compact notation:

$$T_1 = t_1 + r_1, \ T_2 = t_2 + r_2 \text{ and } T_3 = t_3 + r_3. \quad [10, 11, 12]$$

Next we rewrite the above equations for $t_R(k+1)$ as equations for the k-th waiting time $wt_k$.

For sympathetic activation:

$$wt_k = t_R(k+1) - t_R(k) - T_1 \quad [13]$$

For intrinsic activation:

$$wt_k = t_R(k+1) - t_R(k) - T_2 \quad [14]$$

For para-sympathetic activation:

$$wt_k = t_R(k+1) - t_R(k) - T_3 \quad [15]$$

At this point two comments are in order: $[t_R(k+1) - t_R(k)]$ is actually the k-th RR interval henceforth denoted as $RR_k$. Then the waiting time $wt_k$ cannot be negative. Hence, whenever we express $wt_k$ in terms of $RR_k$, the necessary constraints will be in place to enforce this. Further the differential $dwt_k = dRR_k$ as $T_1$, $T_2$ and $T_3$ are constants.

We are now ready to write down the expression for the probability density for $RR_k$ which is:

$$P(RR_1, RR_2, \ldots, RR_N) \prod_{k=1}^{N} RR_k = \prod_{k=1}^{N} \exp(-(\lambda_1 + \lambda_2 + \lambda_3)RR_k)$$

$$\prod_{k=1}^{N} \begin{bmatrix} H(RR_k - T_1)\exp((\lambda_1 + \lambda_2 + \lambda_3)T_1)\lambda_1 \\ +H(RR_k - T_2)\exp((\lambda_1 + \lambda_2 + \lambda_3)T_2)\lambda_2 \\ +H(RR_k - T_3)\exp((\lambda_1 + \lambda_2 + \lambda_3)T_3)\lambda_3 \end{bmatrix}$$

$$\prod_{k=1}^{N} RR_k \qquad [16]$$

To obtain equation 16 we have added the expressions on the RHS of equations 4 to 6, inserted the stated constraints on $wt_k$ through Heaviside step functions H( ), expressed $wt_k$ in terms of $RR_k$ and simplified the resulting expression. The different kinds of activation events are mutually exclusive and hence the addition of the corresponding probabilities is in order.

Appendix I: Monte Carlo implementation of Equation 16

(I) Generate 3 random waiting time sequences corresponding to the sympathetic, intrinsic and para-sympathetic activations. For the standard Poisson point-process these 3 point-processes are fixed by the 3 parameters lambda1 etc. and we invoke 3 exponential random number generators.

(II) Find the least of the three waiting times and the kind of activation from which it arose and add, to this waiting time, the appropriate $T_i$ where i takes the value 1, 2 or 3 depending on the kind of activation. Denote this sum as the current RR interval and also label it appropriately as arising from sympathetic, intrinsic or para-sympathetic activation.

(III) Go to the next three entries in the waiting time sequences and repeat step (II) till as many heart beats as desired have been found. A sample code written in Maple is available with the author.

In fact we may generate, using the above 3 steps, synthetic ECG's for any given values of neural and cardiac data. On the other hand, equation 16 itself provides the necessary multivariate probability density for analytical investigations.

Appendix II: [Method of Extracting Firing Rates from ECG]

In this Appendix we show how we may extract with the help of the proposed model the sympathetic, intrinsic and para-sympathetic tones from real ECG data: The first step will be to compute the RR interval histogram from the ECG data. Then identify the maximum points in the histogram. If there are just 3 points of maximum their values on the abscissa may directly be read as the values of $T_1$, $T_2$ and $T_3$. If the histogram presents less or more than 3 maxima, it can still be analyzed using a simple modification of the model. What we have advanced is an effective model which may have to be tailored to account for features in real-world data.

Once we have obtained the values of $T_1$, $T_2$ and $T_3$ we solve the following set of equations for $\lambda_1, \lambda_2$ and $\lambda_3$:

$\lambda_1$ = Histogram Height at $T_1$ = $H_{T_1}$

$\exp(-(\lambda_1 + \lambda_2 + \lambda_3)T_2) \times \exp((\lambda_1 + \lambda_2 + \lambda_3)T_1)\lambda_1 + \lambda_2 = H_{T_2}$

$\exp(-(\lambda_1 + \lambda_2 + \lambda_3)T_3) \times \left(\exp((\lambda_1 + \lambda_2 + \lambda_3)T_1)\lambda_1 + \exp((\lambda_1 + \lambda_2 + \lambda_3)T_2)\lambda_2\right) + \lambda_3 = H_{T_3}$

A sample computation was performed on a real ECG data taken from the physio-bank ATM and having 24400 heart beats. $T_1$, $T_2$ and $T_3$ were found to be respectively 0.55, 0.63 and 0.72. Solving the above equations after plugging in these values and the value of $H_{T_1}$ leads to the values of 5, 5.481 and 2.159 for $\lambda_1, \lambda_2$ and $\lambda_3$ respectively. Another similar case study produced values of 0.64, 0.82 and 1.08 respectively for $T_1, T_2$ and $T_3$ and 0.01, 1.997 and 5.192 for $\lambda_1, \lambda_2$ and $\lambda_3$ respectively.

**Acknowledgments**

This work is an outcome of an invitation for me to visit IMSc by Professor G. Baskaran using his research grant from SERB Distinguished Fellowship [DST, New Delhi]. He proposed that I work on Heart and I enjoyed several stimulating discussions with him and he brought to my attention a host of research articles relevant to the work reported here. Being new to Heart Research his constant support and encouragement was indeed crucial to me.

Dr. Paul Ramesh, a Distinguished Cardiac-Thoracic surgeon with Apollo Hospital [Chennai] has been playing a very helpful advisory role. And it is further hoped that he and Baskaran will play a very active role in the second phase of this venture which is in its clinical application.

This work was supported by SERB grant no. SB/DF/005/2014 of Department of Science and Technology (New Delhi) awarded to G.Baskaran. Hospitality at IMSc is gratefully acknowledged.